\documentclass{article}
\usepackage{spconf,amsmath,graphicx}
\usepackage{setspace}
\usepackage{hyperref}
\usepackage{siunitx}                
\usepackage{caption}
\usepackage{etoolbox}
\usepackage{tikz}

\newcommand\QP{\mathit{QP}}

\newcommand\copyrighttext{%
	\footnotesize \textcopyright 2020 IEEE. Personal use of this material is permitted.
	Permission from IEEE must be obtained for all other uses, in any current or future 
	media, including reprinting/republishing this material for advertising or promotional 
	purposes, creating new collective works, for resale or redistribution to servers or 
	lists, or reuse of any copyrighted component of this work in other works. 
	DOI: \href{https://doi.org/10.1109/ICASSP43922.2022.9747633}{10.1109/ICASSP43922.2022.9747633} }
\newcommand\copyrightnoticeOwn{%
	\begin{tikzpicture}[remember picture,overlay]
		\node[anchor=north,yshift=-10pt] at (current page.north) {\fbox{\parbox{\dimexpr\textwidth-\fboxsep-\fboxrule\relax}{\copyrighttext}}};
	\end{tikzpicture}%
\vspace{-8mm}
}

\title{Evaluation of Video Coding for Machines without Ground Truth}
\name{Kristian Fischer$^1$, Markus Hofbauer$^2$, Christopher Kuhn$^2$, Eckehard Steinbach$^2$, Andr\'e Kaup$^1$\thanks{The authors gratefully acknowledge that this work has been supported by the Deutsche Forschungsgemeinschaft (DFG) under contract number KA 926/10-1.}}
\address{$^1$Multimedia Communications and Signal Processing, FAU Erlangen-N\"urnberg\\
	$^2$Chair of Media Technology, Technical University of Munich}
\begin{document}
	%
	\maketitle
	\copyrightnoticeOwn
	\begin{abstract}
		In the emerging field of video coding for machines, video datasets with pristine video quality and high-quality annotations are required for a comprehensive evaluation.
		However, existing video datasets with detailed annotations are severely limited in size and video quality.
		Thus, current methods have to either evaluate their codecs on still images or on already compressed data.
		To mitigate this problem, we propose an evaluation method based on pseudo ground-truth data from the field of semantic segmentation to the evaluation of video coding for machines.
		Through extensive evaluation, this paper shows that the proposed ground-truth-agnostic evaluation method results in an acceptable absolute measurement error below 
		0.7 percentage points on the Bj\o ntegaard Delta Rate compared to using the true ground truth for mid-range bitrates. We evaluate on the three tasks of semantic segmentation, instance segmentation, and object detection.
		Lastly, we utilize the ground-truth-agnostic method to measure the coding performances of the  VVC compared against HEVC on the Cityscapes sequences. This reveals that the coding position has a significant influence on the task performance.
	\end{abstract}
	\vspace{-2mm}
	\section{Introduction}
	\label{sec:Introduction}
	\vspace{-1mm}
	Thanks to a drastic increase in accuracy and applicability, a significant amount of computer vision tasks is solved by neural networks today, such as in autonomous driving or surveillance of public spaces. 
	In such applications, machines are directly interacting with each other without a human being involved.
	Hence, this type of communication is typically referred to as machine-to-machine~(M2M) communication.
	Cisco underlines the relevance of this emerging topic by stating that half of all global devices and connections will be accounted to M2M communication in 2023~\cite{cisco2020}.
	
	In practical applications, the input multimedia data are usually compressed and transmitted before being analyzed by the machine.
	This saves eventual transmission rate or storage space on hard disks.
	Together with the growing importance of M2M communication, this requires specialized compression schemes in M2M applications.
	Early work in this area optimized the compression to preserve extracted features~\cite{chao2011preserving,chao2012sift,chao2013design}.
	More recent work focuses on coding for M2M communication, where a neural network analyzes the decoded image instead of a human observer~\cite{choi2018,galteri2018,fischer2020_ICIP,fischer2020_FRDO, le2021_ICASSP, fischer2021_ICASSP}.
	This topic is commonly referred to as video coding for machines~(VCM).
	Additionally, the Moving Picture Experts Group (MPEG) founded an ad hoc group on VCM in 2019, with the objective of standardizing an optimal bitstream for M2M communication~\cite{zhang2019}.
	
	To evaluate codecs for a VCM scenario, an arbitrary algorithm or neural network is applied to the compressed input image and its performance is measured depending on the deteriorations present.
	Thus, there are three main conditions for a proper dataset:
	1.)~The dataset needs to contain uncompressed, pristine input data, such that the coding efficiency measurement is not falsified by already existing artifacts.
	2.)~The dataset requires properly labeled ground-truth~(GT) data to measure the performance of the applied algorithm.
	3.)~The dataset needs to contain video data on top/instead of images, since practical scenarios are usually based on video streams.
	To the best of our knowledge, there is currently no dataset available that fulfills all three conditions.
	Thus, previous VCM research either sacrifices the condition of uncompressed input data~\cite{choi2018, galteri2018} or evaluates the coding frameworks on single images~\cite{fischer2020_ICIP, fischer2020_FRDO, le2021_ICASSP, fischer2021_ICASSP}.
	
	\begin{figure*}[t]
		\centering
		\includegraphics[width=0.9\textwidth]{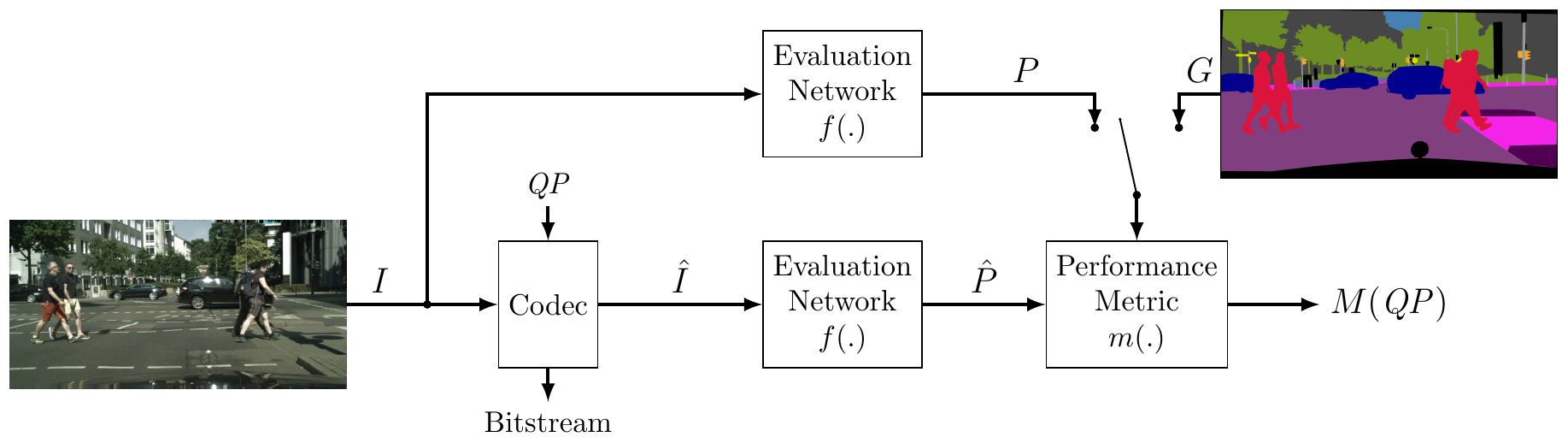}
		\caption{VCM evaluation framework. The switch either selects the traditional evaluation framework with true GT data $G$, or the GT-agnostic evaluation with the pseudo GT data $P$.}
		\label{fig:vcm framework}
		\vspace{-6mm}
	\end{figure*}
	
	To alleviate the problem of requiring hand-labeled annotations, which most datasets do not have, several papers from the research field of future semantic segmentation either turn to simulated data~\cite{kuhn2021pixel} or generate pseudo GT data. Since we focus on real world applications, we do not consider simulation-based approaches. Luc~et~al.~\cite{luc2017} and Nabavi~et~al.~\cite{nabavi2018FutureSS} take the predictions of a state-of-the-art semantic segmentation network as pseudo GT to evaluate their future semantic segmentation models.
	Couprie~et~al.~\cite{couprie2018} applied pseudo GT annotations for the task of future instance segmentation for the Cityscapes~\cite{cordts2016} dataset.
	Aqqa~et~al.~\cite{aqqa2019} already measured the VCM performance of four different state-of-the-art object detection networks on different H.264/AVC~\cite{wiegand_h264} compression levels taking the predictions on uncompressed data as GT.
	However, they do not provide any statement on how accurate their measurement is compared to GT-based evaluations.
	
	Inspired by those approaches, this paper proposes to apply a GT-agnostic evaluation framework to evaluate VCM scenarios.
	We use the network predictions on uncompressed, pristine input data as pseudo GT.
	Thereby, this paper contributes an extensive analysis of the GT-agnostic evaluation framework and quantifies the measurement error compared with the traditional GT-based method.
	Second, with the help of the GT-agnostic method, we compare the two latest video codecs High Efficiency Video Coding~(H.265/HEVC)~\cite{sullivan2012_HEVC} and its successor Versatile Video Coding~(H.266/VVC)~\cite{chen2020vtm10} for inter coding in \textit{randomaccess} configuration on the non-labeled Cityscapes sequences.

	\section{Analyzing VCM with GT-Agnostic Evaluation}
	
	\subsection{Traditional VCM Evaluation with True GT}
	
	In a typical VCM scenario for image coding as depicted in \autoref{fig:vcm framework}, the input image $I$ is first encoded into a compact bitstream representation, which is steered by the user-defined quantization parameter~($\QP$) towards either a low bitrate or a low distortion.
	Afterwards, the encoded bitstream is decoded resulting in the deteriorated output image $\hat{I}$.
	In classic compression frameworks focusing on human perception, $\hat{I}$ is compared against $I$ to calculate distortion metrics representing the human visual system.
	In VCM however, $\hat{I}$ is fed into an arbitrary evaluation network $f(.)$, which fulfills a certain task resulting in the predictions $\hat{P}$.
	Subsequently, those predictions are compared against the ground-truth data $G$ by a performance metric $m$ that measures the performance of the evaluation network on the distorted input. The resulting performance score $M$ can be formulated as
	\begin{equation}
		M(\QP) = m(\hat{P}(QP), G) = m(f(\hat{I}(\QP)), G).
	\end{equation}
	
	We investigate the scenarios of semantic and instance detection/segmentation, using the mean intersection over union~(mIoU) and the average precision~(AP)~\cite{hariharan2014_AP_original} as performance metrics, respectively.
	Both return values between 0 and 1, with 0 being not accurate and 1 representing perfectly matched GT instances by the neural network.
	Ultimately, we plot the resulting performance $M(\QP)$ over the required bitrate.
	This indicates how well the neural network performs on compressed input images.
	
	\subsection{Proposed GT-Agnostic Evaluation with Pseudo GT}
	\label{subsec:Gt-agnostic vcm evaluation}
	
	Without the presence of properly annotated GT data $G$, we propose that pseudo GT labels are sufficient to evaluate the coding efficiency in VCM scenarios.
	As depicted in \autoref{fig:vcm framework}, we consider the predictions $P$ of the evaluation network on the pristine input data $I$ as pseudo GT for our GT-agnostic evaluation framework:
	\begin{equation}
		G_{pseudo} = P = f(I).
	\end{equation}
	With that, the overall performance measurement changes to
	\begin{equation}
		M_{pseudo}(\QP) = m(f(\hat{I}(\QP)), G_{pseudo}).
	\end{equation}
	
	The predictions $P$ can be considered as pseudo GT data for the predictions $\hat{P}$ derived from the compressed inputs.
	It is shown in the VCM-related literature as well as later in Sec.~\ref{subsec:Analysis of GT-agnostic VCM Framework} that the predictions on compressed inputs have a lower performance than the pseudo-GT.
	This is similar to future instance segmentation, where the network segmenting the current frame has a higher performance than the networks predicting the segmentation masks based on the past.
	
	In general, $M(\QP)$ and $M_{pseudo}(\QP)$ differ significantly.
	For the metric result $M(\QP)$ measured on true GT, the maximum possible value is below 1, since real-world networks are typically not perfectly matching the GT.
	For the pseudo-GT based performance $M_{pseudo}(\QP)$, this maximum is 1 when the predictions $\hat{P}$ on the compressed inputs are equal to the predictions on pristine data $P$.
	
	However, the leading metric for comparing the coding quality of two codecs regarding VCM is the Bj\o ntegaard Delta Rate~(BDR)~\cite{bjontegaard2001_new}, such as used in \cite{choi2018, fischer2020_ICIP, le2021_ICASSP}.
	For VCM, it defines how much rate is saved while keeping the performance of the network the same for a certain quality range.
	Therefore, the absolute values of $M(\QP)$ and $M_{pseudo}(\QP)$ are not important, but rather their relative behavior over the required bitrate.
	
	\section{Analytical Methods}
	\label{sec:Analytical Methods}
	
	\begin{figure}[t]{}
		\centering
		\includegraphics[width=\columnwidth]{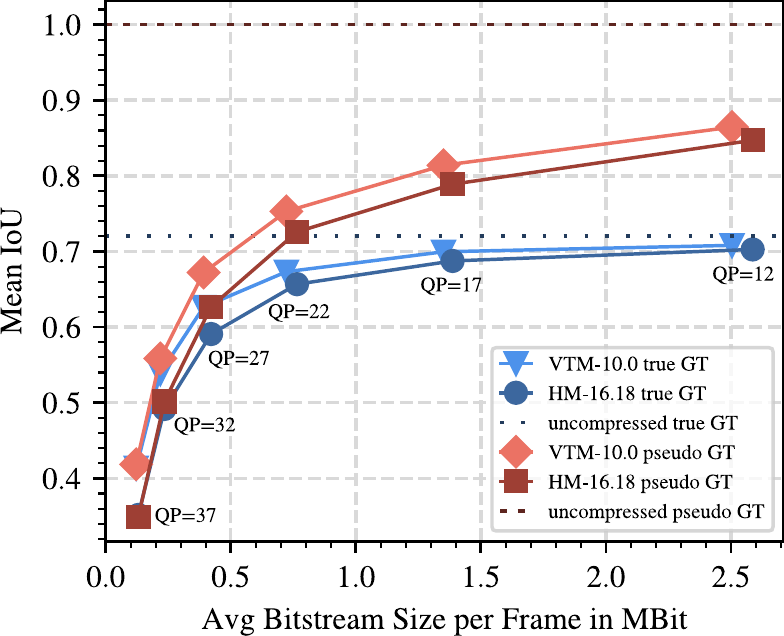}
		\caption{Mean IoU measured with true GT (blue) and pseudo GT (red) for DeepLabV3+ over the bitrate when coding with VTM-10.0 and $\QP \in \{12,17,22,27,32,37\}$ for the 500 Cityscapes validation images. The dotted lines represent the corresponding mean IoU on uncompressed input images.}
		\label{fig:performance ap over bitrate}
		\vspace{-4mm}
	\end{figure}
	
	To quantify the measurement error of the GT-agnostic VCM evaluation over the traditional evaluation method, we select the Cityscapes validation dataset comprising 500 pristine and annotated images with a spatial resolution of $1024\times2048$ pixels.
	We compress these images with the two standard-compliant codecs HEVC test model~(HM-16.18)~\cite{hm_software} and VVC test model~(VTM-10.0)~\cite{vtm_software} for two different $\QP$ ranges.
	The first set of $\QP$ is $\QP\in\{22, 27, 32, 37\}$ as defined in the JVET common testing conditions~(CTC)~\cite{bossen2020_VVC_CTC}.
	Additionally, we test higher bitrate and quality ranges for $\QP \in \{12, 17, 22, 27\}$ to obtain values that are closer to the neural network performance on pristine data similar to the investigations in~\cite{fischer2020_FRDO}.
	To represent the human visual system, we also consider PSNR and VMAF~\cite{li2016_vmaf_short}.
	
	As an evaluation network for the task of semantic image segmentation, we employ the state-of-the-art model DeepLabV3+~\cite{chen2018_DeepLabV3+} with a Mobilenet~\cite{sandler2018_mobilenet} backbone and the pre-trained models from~\cite{deeplabRepo}.
	For instance segmentation, we use Mask~R-CNN~\cite{he2017} with the weights from~\cite{wu2019detectron2}.
	For object detection with Faster R-CNN, we trained a model similar to~\cite{fischer2020_ICIP} on Cityscapes.
	Both R-CNN networks have a ResNet-50 backbone with a feature-pyramid structure~\cite{lin2017fpn}.
	
	We measure the performance of the semantic segmentation using the mIoU, which is the standard metric for evaluating on the Cityscapes dataset~\cite{cordts2016}.
	For instance segmentation, we measure the weighted AP~(wAP), which weights the AP for the eight Cityscapes classes depending on the frequency of their occurrence as we have proposed earlier in~\cite{fischer2020_ICIP}.
	
	\section{Experimental Results}
	\subsection{Analysis of GT-agnostic VCM Framework}
	\label{subsec:Analysis of GT-agnostic VCM Framework}
	
	\begin{table}
		\centering
		\normalsize
		\caption{BDR values of VTM over HM in \si{\percent} with the corresponding quality or performance metric as reference for the 500 labeled Cityscapes validation images. \\
			$\QP \in \{12,17,22,27\}$.}
		\vspace{-2mm}
		\begin{tabular}{l|rrr}
			\hline
			& True GT     & Pseudo GT & Diff. \\
			\hline
			PSNR    & \num{-16.62}    & -    & -                      \\
			VMAF    & \num{-30.51}    & -    & -                        \\
			mIoU DeepLabV3+        & \num{-30.06} & \num{-25.25}   & \num{-4.81}           \\
			oAcc DeepLabV3+    & \num{-40.04} & \num{-34.37}   & \num{-5.67}           \\
			frwAcc DeepLabV3+      & \num{-39.69} & \num{-34.07}   & \num{-5.62}           \\
			wAP Mask R-CNN   & \num{-7.48}  & \num{-12.34}   & \num{4.86}            \\
			wAP Faster R-CNN & \num{-9.00}  & \num{-12.38}   & \num{3.38}            \\
			\hline
		\end{tabular}
		\label{tab:BDR values theta and delta QP low QP}
	\end{table}
	
	\begin{table}[]
		\centering
		\normalsize
		\caption{BDR values of VTM over HM in \si{\percent} with the corresponding quality or performance metric as reference for the 500 labeled Cityscapes validation images.\\$\QP \in \{22,27,32,37\}$.}
		\vspace{-2mm}
		\begin{tabular}{l|rrr}
			\hline
			& True GT     & Pseudo GT & Diff. \\
			\hline
			PSNR             & \num{-22.74} & -   & -             \\
			VMAF             & \num{-26.01} & -   & -              \\
			mIoU DeepLabV3+        & \num{-28.31} & \num{-27.62}   & \num{-0.69}          \\
			oAcc DeepLabV3+     & \num{-43.62} & \num{-42.96}   & \num{-0.66}           \\
			frwAcc DeepLabV3+       & \num{-43.49} & \num{-43.52}   & \num{0.03}            \\
			wAP Mask R-CNN   & \num{-13.30} & \num{-13.06}   & \num{-0.24}            \\
			wAP Faster R-CNN & \num{-10.66} & \num{-11.22}   & \num{0.56}            \\
			\hline
		\end{tabular}
		\label{tab:BDR values theta and delta QP CTC}
		\vspace{-3mm}
	\end{table}
	
	In \autoref{fig:performance ap over bitrate}, we draw the rate-performance curves of HM and VTM coding the Cityscapes validation dataset depending on whether the mIoU for semantic segmentation has been measured with or without true GT.
	The blue colors represent the traditional performance measurement with true GT data, whereas the red colors represent the GT-agnostic measurement.
	There, the differences between both evaluation approaches regarding absolute mIoU values can be clearly seen.
	As mentioned in Sec.~\ref{subsec:Gt-agnostic vcm evaluation}, the performance for the uncompressed GT-agnostic case is 1.
	For the traditional method, the performance for a low $\QP$ value of 12 is almost equal to the equivalent uncompressed case.
	For pseudo-GT, the performance drops drastically by around 15 percentage points when using input data compressed with $QP=12$ instead of pristine input images.
	A possible explanation for this behavior is that the applied DeepLabV3+ network is not perfect, as indicated by the mIoU value of around \SI{70}{\percent} evaluated on the true GT data.
	These model imperfections make the IoU measurement more sensitive to small noise-like differences between the uncompressed and the slightly compressed data, which ultimately results in this large drop.
	Considering the four highest $\QP$ values, the influence of the noise declines and the basic behavior between HM and VTM is very similar for both measurement methods.
	
	However, as mentioned in Sec.~\ref{subsec:Gt-agnostic vcm evaluation}, the evaluation of VCM is focused on the relative differences between the different codecs regarding BDR instead of the absolute performance values.
	Thus, we calculate the BDR values of VTM over HM and list them in \autoref{tab:BDR values theta and delta QP low QP} for high bitrates and in \autoref{tab:BDR values theta and delta QP CTC} for the CTC-recommended values.
	This confirms the visual impression from \autoref{fig:performance ap over bitrate}.
	For low $\QP$ values, the difference between measuring the BDR with or without the GT data is \num{4.81} percentage points for the semantic segmentation with DeepLabV3+ regarding mIoU.
	For the JVET-recommended bitrate ranges, the total difference between calculating the BDR with true GT or pseudo GT reduces to only \num{0.69} percentage points.
	
	\begin{table}
		\centering
		\normalsize
		\caption{BDR values of VTM over HM in \si{\percent} and \textit{randomaccess} configuration for the corresponding quality or performance metric using proposed pseudo GT for the Cityscapes validation sequences and $\QP \in \{22,27,32,37\}$.}
		\begin{tabular}{l|r}
			\hline
			& BDR \\
			\hline
			PSNR             & \num{-33.19}               \\
			VMAF             & \num{-38.34}                \\
			mIoU DeepLabV3+       & \num{-33.62}        \\
			wAP Mask R-CNN   & \num{-21.04}           \\
			wAP Faster R-CNN & \num{-20.86}           \\
			\hline
		\end{tabular}
		\label{tab:BDR values inter coding}
		\vspace{-3mm}
	\end{table}
	
	We additionally investigated other standard semantic segmentation metrics such as the overall accuracy~(oAcc) and frequency weighted accuracy~(frwAcc), resulting in similar differences between both evaluation methods.
	Besides, those values are also confirmed for the task of instance segmentation with Mask R-CNN and object detection with Faster R-CNN, with absolute errors of \num{0.24} percentage points and \num{0.56} percentage points, respectively.
	This shows that the GT-agnostic measurement is a well-suited tool for evaluating VCM scenarios without having hand-labeled GT data for the JVET-CTC recommended QP range.
	
	
	\vspace{-1mm}
	\subsection{Performance HEVC vs. VVC for Inter Coding}
	\label{subsec:Performance HEVC vs. VVC for Inter Coding}
	
	After demonstrating that the GT-agnostic evaluation method performs quite on par with the traditional one, we show the coding gains of VTM over HM for coding Cityscapes videos.
	For our evaluations, we compressed the frames 16 to 24 of a Cityscapes validation sequence with the originally labeled $20^\mathrm{th}$ frame of one Cityscapes sequence being placed in the middle.
	The chosen codec configuration was \textit{randomaccess}.
	In \autoref{tab:BDR values inter coding} we present the BDR values for the different distortion and performance metrics. The coding gains for VTM over HM are similar when coding for DeepLabV3+ compared to using PSNR and VMAF. For the tasks of instance segmentation and detection however, the coding gains of VTM are significantly lower compared to the other metrics. These values show the same tendency as in the intra coding case listed in ~\autoref{tab:BDR values theta and delta QP CTC} and extensively discussed in~\cite{fischer2020_ICIP}. 
	
	\begin{figure}[t]
		\centering
		\includegraphics[width=0.405\textwidth]{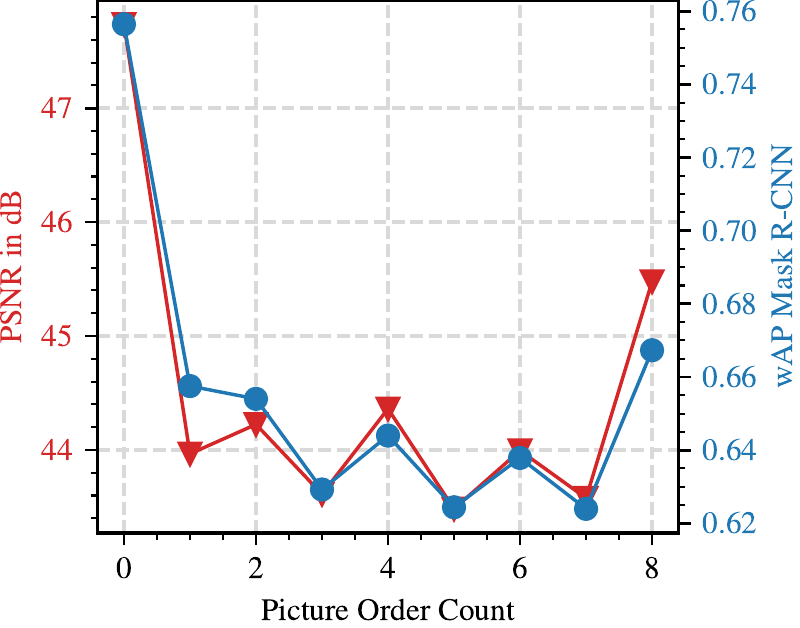}
		\caption{PSNR (red) and pseudo-GT-based wAP (blue) for Mask R-CNN over the picture order count averaged over all sequences compressed with VTM-10.0 and $\QP=22$.}
		\label{fig:performance over POC}
		\vspace{-3mm}
	\end{figure}
	
	When coding videos for neural networks in practical applications, it is important that the detection performance is constant regardless of the position were the frame has been coded.
	However, as drawn in \autoref{fig:performance over POC}, the detection quality of Mask R-CNN strongly fluctuates depending on when the frame is coded in the \textit{randomaccess} order, with the highest detection performance for the I-frame.
	This indicates that \textit{randomaccess} coding might be infeasible for certain practical VCM applications.
	\vspace{-1mm}
	\section{Conclusion}
	
	In this paper, we propose to evaluate VCM scenarios by employing pseudo-GT data derived from the predictions of computer vision models trained on uncompressed multimedia data.
	A broad analysis for the $\QP$ values defined in the JVET CTCs and three different use cases on the Cityscapes dataset revealed that measuring the coding performance between HM and VTM with this GT-agnostic method only results in negligible errors of up to \SI{0.69}{\percent} in the worst case compared with the measurement using true GT data.
	Thus, our work suggests that future VCM evaluations can also be done by considering uncompressed video datasets without hand-labeled annotations such as the HEVC test sequences or self-captured datasets in practical applications, where annotating is commonly a cumbersome and costly process. The main limitation of this proposed evaluation scheme is that suitable machine-learning tasks and models have to be defined beforehand.
	
	Finally, applying the GT-agnostic method to evaluate \textit{randomaccess} coding on the Cityscapes sequences showed that the coding gains for instance segmentation and detection are not as high as for human-visual-based metrics. 
	It has to be further researched whether the fluctuating detection performance depending on the coding order is negatively affecting practical applications. This could be of special interest for tasks requiring video data as input such as tracking.
	
	\vfill\pagebreak
	\bibliographystyle{IEEEbib}
	\setstretch{0.5}
	\small
	\bibliography{literature.bib}
\end{document}